
\magnification=1200
\vskip 3.cm
\centerline {\bf MEAN-FIELD DESCRIPTION OF NUCLEAR}
\centerline {\bf CHARGE DENSITY DISTRIBUTIONS}
\vskip 2.cm
\centerline {R.Anni and G.Co'}
\vskip .4cm
\centerline {Dipartimento di Fisica, Universit\`a di Lecce}
\centerline {and}
\centerline {INFN, Sezione di Lecce}
\centerline {I-73100 Lecce, Italy}
\vskip 1cm

{\bf Abstract.} \quad
We propose a method to extract nuclear charge distributions from
elastic electron scattering data based upon a mean field approach.
The nuclear charge distributions are generated by solving the
Schr\"odinger equation with a  mean-field potential expanded in
terms of Hermite functions. The coefficients of the
potential are changed in order to obtain the best fit
with the cross section data.
Applications to the $^{12}C$, $^{16}O$ $^{40}Ca$ and $^{208}Pb$
nuclei are presented.

\vfill\eject

{\bf 1. Introduction}
\vskip 0.5 cm

The distribution of the electromagnetic charge in the nuclear
ground state is one of the more interesting, and investigated,
properties of the atomic nuclei. Nowadays, the widest and most
accurate set of information about this quantity is provided by
elastic electron scattering experiments. Ultra relativistic
electrons are scattered by the electrostatic potential generated by
the nuclear charge distribution. From the measured cross section one
extracts information about this distribution.

The straightforward approach to get the charge distribution is the
solution of the inverse scattering problem. Unfortunately
there are principle and pragmatical difficulties making
this approach extremely impractical [1] and for this reason it has
never been applied in this context.

The approach which has been commonly used consists in solving
directly the scattering problem making a guess about the form of the
potential. The parameters fixing the form of the potential are
changed until the experimental cross sections are
reproduced.

In early times, the quality of the data available was such that
they could be fitted reasonably well using simple functions
containing few parameters (e.g. Fermi distributions).
In the last two decades, thanks to the the development of
experimental facilities, there has been a big
improvement of both quality and quantity of the data.
Reasonable fits of modern elastic cross section data can be obtained
only increasing the number of the free parameters.
Instead of
inserting a dependence of new parameters in the old functions
describing the densities it results more convenient to make use of the
so-called model independent techniques [2,3].

These techniques consist in expanding the charge
distribution on a orthonormal basis and changing the coefficients of
the expansion to achieve the best fit of the cross section.
In principle the model independent techniques are very general and
clean, but their application to the specific problem of fitting
experimental cross sections shows ambiguities and uncertainties.
The expansion of the density on a basis does not converge
but rather, with the increase of the number of the expansion terms
shows an enhancement of the uncertainty band of the charge
distribution. All this happens in spite of the fact that the quality
of the fit to the cross section remains constant.
We have shown in ref. [4] that this instability of the method is
not related to a specific choice of the expansion basis, but it is
rather a general drawback of the model independent techniques.

Pragmatical recipes have been developed to keep under control the
instabilities of the model independent techniques. These are ranging
from analytical continuations of the experimental data in the
unmeasured region [3] to the identification of the optimal number
of expansion coefficients [4].

This unsatisfactory situation still reflects the fact that it is not
possible to solve the inverse scattering problem for a set of data
known on a discrete and finite grid and with statistical error [1].

In the present article we propose an extraction method
alternative to the model independent techniques. This method is in
principle less general, but it produces more stable solutions.

Our idea is to use the mean field model of the nucleus to extract
the charge distribution from the experimental data.
The nucleus is described in the framework of a conventional shell
model. The charge distribution is given by the convolution of the
nucleon electromagnetic form factor with nucleon densities generated
by the sum of the squares of the single particle wave functions. The
parameters describing the nuclear average potential generating the
wave functions can be changed to fit the cross section data.

It is clear that this approach has more physical insight than the
model independent techniques, and it is restricting the class of
functions used for the description of the charge distributions. For
example in this case the charge distribution is ensured to be always
positive; this is not the case for the model independent techniques.

For different purposes, fits of elastic
electron scattering data
using the nuclear shell model with
an energy dependent mean--field,
have been published
already in the 1967 [5].
The basic idea of this work was to use
the experimental data in order to fix an empirically well grounded
mean--field potential.

For us the shell model is only a tool which imposes
reasonable physics constraints on the class of functions used to
build up the charge distributions.
We are showing that energy independent mean--fields with enough
degrees of freedom can reproduce elastic electron scattering data on
a much wider momentum transfer range than that available in 1967.
In spite of this we are not attributing any special meaning to the
potentials obtained.

In addition, we do not consider the charge
distributions we present in this work as the most realistic ones.
We are conscious that a careful extraction of the nuclear charge
distribution should take care that the scattering data are affected
by dispersion effects, meson exchange currents, neutron charge
distribution and other phenomena which are ignored in this work.

The aim of the present paper is to present an extraction
method providing charge distributions which result to be
much more stable against the number of the free
coefficients than the traditional model independent techniques.

In the next section we describe the method used to generate the
mean field potential and we test its performances on pseudodata. The
method is applied in sect. 3 to reproduce the experimental
cross section of the $^{12}C$, $^{16}$O, $^{40}Ca$ and $^{208}Pb$
nuclei.

\vskip 1 cm
{\bf 2. A generalized mean-field model.}
\vskip 0.5 cm

As discussed in the introduction, we are working in the framework of
the direct
scattering approach. We build up a charge distribution $\rho_c(r)$
generating the electrostatic potential used to solve the Dirac
equation for ultra-relativistic electrons.
The techniques used to solve the Dirac equation, and to obtain the
elastic scattering cross section,  are presented in ref. [6].

The charge distribution is obtained folding the nucleon density
distribution $\rho({\bf r})$ with the nucleon form factor
$f_n({\bf r}',{\bf r})$:
$$
\rho_c({\bf r})=\int \rho({\bf r'}) f_n({\bf r}',{\bf r }) d^3 {\bf
r'}  \eqno (2.1)
$$

The density distribution of a system of $A$ particles is defined as:
$$
\rho({\bf r} )= <\Psi | \sum_{i=1}^A \delta ({\bf r} - {\bf r_i})
|\Psi>  ,
\eqno(2.2)
$$
where we have indicated with $|\Psi>$ the wave function describing the
ground state of the system.

In the mean field approximation $|\Psi>$ is a Slater determinant
of single particle wave functions $\phi_\alpha({\bf r})$
and the density can be written
as:
$$
\rho({\bf r})=\sum_\alpha \phi_\alpha({\bf r})^*
\phi_\alpha({\bf r}),
\eqno(2.3)
$$

where $\alpha$ indicates the set of quantum numbers identifying the
single particle level.

We work with a mean field potential of the form:

$$ V_{MF}(r)=V(r)+ {\lambda \over r}  {d \over dr}  V(r) \;
{\bf l} \cdot {\bf s} + V_{Coul}(r) \eqno(2.4)
$$

where $V_{Coul}$ indicates the Coulomb potential which has been
chosen as the electrostatic potential generated by a uniformly
charge distribution of radius $1.2 A^{1 \over 3}$ fm.

In this expression we should fix the expression of the central
term $V(r)$ and the strength $\lambda$ of the spin-orbit term.

Our method consists in
expanding $V(r)$ on a basis of orthonormal functions $p_n(r)$:
$$
V(r)= \sum_{n=0} ^\infty c_n p_n(r) .
\eqno(2.5)
$$

Specifically
we have chosen $p_n(r) =$ $u_n( r / \beta)$ where the $u_n$ are
Hermite functions defined as [6]:
$$
u_n(x) = ( \sqrt \pi 2^n n! )^{ -{1 \over 2} }
e^{- {x^2 \over 2} }
H_n(x) ,
\eqno(2.6)
$$
where we have indicated with $H_n$ the Hermite
polynomial of  order $n$.
The orthogonality relation of the $u_n$ functions is
given by:
$$
\int_{-\infty}^{\infty}
u_n(x) u_m(x) \, dx = \delta_{n,m}.
\eqno(2.7)
$$

The calculations we have performed consist in solving the
single-particle Schr\"odinger equation with the potential given by
the eqs. (2.4) and (2.5). With the single particle wave functions
obtained in this way we construct the density distribution, eq.
(2.3), and then the charge distribution, eq. (2.1).
With this charge distribution we build up the Coulomb potential to
solve the Dirac equation and to obtain the cross section which is
compared with the experimental data. The procedure is repeated
modifying the parameters $c_n$ of the expansion of the potential
until the minimum value of the $\chi^2$ is reached.

As a general rule we have populated the single particle levels
which are usually thought to be occupied in the conventional shell
model description of nuclear ground states.

We have tested the procedure described above applying the method to
reproduce $^{208}$Pb pseudodata generated by a mean-field model with
Woods-Saxon potential. The pseudodata have been obtained
substituting in eq. (2.4):
$$
V(r)={ -V_0 \over 1+exp({ r-R \over a}) }. \eqno(2.8)
$$
with the parameters given in Tab.1.

We have tested the sensitivity of our results to the choice of the
nucleon form factor, comparing charge distributions and cross
sections calculated with the same single particle wave functions but
with various form factors. The differences between the various
results are really negligible, contrary to what happens in the
quasi-elastic peak region [8].  We have
used the form factors given in the reference [9], but
neglecting the small contribution of the neutrons. This means that
the sum in equation (2.3) runs only on the protons.

The charge distribution generated by the Woods-Saxon potential
has been used to produce cross section pseudodata under the same
kinematical conditions of the $^{208}Pb$ experimental data of
ref.[10,11]. The percentage errors on the cross section
pseudodata have been taken equal to the experimental ones [10,11].

We found that the straightforward application of our minimization
techniques, based on the gradient method, had
convergence problems. This was due to the fact during the variation
of the parameters, some of the populated
single particle levels go in the continuum, producing
discontinuities in the charge distribution and therefore in the
$\chi^2$.

In order to force the parameters to move only in regions in which all
the conventional shell model single particle states are bound, we
applied the gradient minimization  technique to the sum of the
$\chi^2$ of the cross section and of the  $\chi^2$ of the single
particle levels.

To the above sum we added a penalty term:
$$
P=\mu \sum_i V(r_i)^2 ,\eqno(2.9)
$$
where the index $i$ runs over all the points of the grid used in the
numerical  integration of the Schr\"odinger equation where
$V(r_i)>0$.

We have added this term to insert a reasonable physics input in
order to speed up the minimisation procedure guiding the algorithm
towards regions, in the space of the parameters, physically
meaningful. This is the analogue in a model independent techniques of
including a penalty for negative charge distributions. In all the
fits performed, the final value of the
penalty term resulted to be compatible with zero.

In this way the merit function which is minimized depends on the
strength of the penalty parameter $\mu$ and on the
errors entering in the $\chi^2$ associated to
the single particles levels.
Simple recipes fixing these
quantities do not exist. Our experience indicates that
only repeated minimization performed by varying the  relative
importance of the terms of our merit function produce
satisfactorily results.

Using these prescriptions we obtained excellent fits to both the
cross section and the single particle energies.

Some results of the fit to the pseudodata are shown in fig. 1.
In the upper panel we compare the density obtained by
the fit to the cross section with 8 Hermite parameters, with the
original Woods-Saxon density shown by the full line.
On the scale of the figure the lines are overlapped.
In the other two panels of the figure
we compare the central and the spin-orbit part of the Woods-Saxon
potential (full lines) with the potential generated by the Hermite
expansion given in eq (2.4) (dashed lines).

The central part of the potential is rather well reproduced,
especially on the surface region. In the internal region the
potential generated by the Hermite expansion shows small
oscillations.

While the central potentials are rather similar, the spin-orbit
potentials show big differences. The spin-orbit potential generated
by the Hermite expansion (dashed line) has big oscillations in the
interior of the nucleus (for values of the radius smaller than 5 fm).
This behavior is not present in the Woods-Saxon spin-orbit term,
which, in this region, is smoothly going to zero. On the other hand,
the two potentials are rather similar in the surface region. This
behavior is not surprising but it comes out from the fact that we
chose to determine the spin-orbit potential from the derivative of
the central term [12]. As we have discussed before the
central term are similar on the surface but they differ in the
interior by the presence of oscillations which are generating the
curious behavior of the spin-orbit term in this region.

To get rid of the oscillations in the internal region of the the
spin-orbit term, we have repeated the fit constructing this
term in a different manner. We calculate the derivative of the
central potential starting from the external part of the nucleus,
and after the first maximum of the derivative we build up the rest of
the spin-orbit potential as symmetrical function with respect to
this point. The dashed-dotted lines of fig. 2 have been generated
with this method.

The results obtained in this way are very similar to the previous
ones. This tells that the internal part of the spin-orbit potential
has very little influence on the results. This fact can be
understood considering the fact that only the s-waves are sensitive
to the details of the potential in the interior of the nucleus. On
the other hand the spin-orbit potential does not act on these waves.

In the model independent procedures the increase of the
number of expansion terms is producing a widening of the uncertainty
band of the charge distribution. We have tested the stability of our
model against the increase of the number of the expansion terms
repeating the fit to the cross section with different number of
expansion terms. The results of this study are shown in fig.
2. One can see that there is a remarkable stability of charge
distribution. This stability of the charge distribution does not
correspond to a stability of the potential, which for a large number
of coefficients starts to have an oscillating behavior.

These oscillations are produced by the high components of the
expansion (2.5). This effect is rather well known in the model
independent extraction of the density [2,3,4] and it is the
origin of all the problems connected with the increasing of the
uncertainty of the charge distribution.
What fig. 2 is showing it is
one of the advantages of performing a fit of the cross section using a
mean field approach instead than a direct expansion
of the density. The high components of the expansion are
acting on the
potential, but the density, generated by the sum of the single
particle wave functions, remains stable.

\vskip 1 cm
{\bf 3. Application to specific cases.}
\vskip 0.5 cm

The results presented
in the previous section show that the mean field method
we have proposed is able to reproduce the set of pseudodata generated
by a mean field potential with great accuracy.
We do not expect the same kind of performances when the
method is applied to fit the experimental data.

We should remark that the model independent expansion techniques of
the density produce fits to the pseudodata of the same quality of that
obtained by our mean field approach.

In this section we present the results we have obtained using the
mean field method to reproduce the electron
scattering elastic cross sections measured on the $^{12}C$,
$^{16}O$, $^{40}Ca$ and $^{208}Pb$ nuclei [10,11,13,14,15].

As we have discussed in the case of pseudodata, our minimization
procedure needs the knowledge of the single particle energies. The
value of the experimental single particle energies we have used are
presented in tab. 2 together with our results.  We remark that the
use of the levels presented in tab. 2 is sufficient to stabilize the
minimization procedure.

The agreement between the calculated energies and
experimental ones is reasonable,
but not as good as in the case of the pseudodata.
The experimental values of the single particle
energies we have used have been obtained with different methods
(extracted from the binding energies of neighbor nuclei, from the
centroid energies in knock-out experiments, etc.).
We are identifying these energy values with the eigenvalues of the
single particle hamiltonians building up the mean field many body
hamiltonian.  This identification is quite extreme and doubtful.
Many-body effects present both in the ground state properties and in
the extraction mechanism can strongly modify the interpretation in
terms of mean field approach.

A comparison of the performances of our approach against those of
the model independent techniques is shown in fig. 3 where the charge
distributions obtained fitting the $^{40}Ca$ data [11] with a
Fourier--Bessel (FB) expansion of the density  and with our method
are shown.

The dark areas represent the region defined by the upper and lower
envelopes of the densities and potentials obtained by a
set of 100 fits of the experimental cross section normally sampled
within the error band [4]. The FB densities,
shown in the panels A and B, have been obtained with 9 and 13
coefficients respectively. In the panels C and D we show the
densities obtained with our model and the mean field
potentials (right vertical scales and lighter areas).
In C the fit has been done with 8 expansion coefficients, while in D
with 14 coefficients.

The distribution bands of the densities and the averaged reduced
$\chi^2$ show that the quality of the fits obtained with our model is
analogous to that obtained by the usual model independent
techniques in their better performances.
In the FB case the increase of the
number of free parameters generates large charge distribution
bands. With our method the instability
generated by increasing the number of the expansion coefficients
shows up in the potential, but the charge distributions remain
quite stable.

In fig. 4 we compare the results of our method with those
obtained using the FB expansion of the density
for all the nuclei we have considered. In this figure the full
lines have been obtained with the shell model,
while the dispersion band has been obtained by a set of FB
fits. As we have seen in fig. 3 the dispersion band of our
model is of the same order of magnitude of that of the FB fits.

The full lines are completely overlapped by the FB density bands.
Only in the central region of $^{16}O$ we notice some difference
between the FB density band and the full line produced by the shell
model fit.

In tab. 3 we
compare the values of the reduced $\chi^2$ obtained for various
fits. We observe that the mean field fits have $\chi^2$ values
comparable to those obtained with the FB expansion and slightly
larger than those obtained with the Hermite expansion.

We do not insert figures of the cross sections, since on the usual
printable scale, the experimental data and all the calculated lines
are overlapping, as clearly indicated by the values of
the $\chi^2$ of tab.3.

The potentials generating the single particle wavefunctions which
produce the densities fitting the experimental data are shown in
fig. 5. We observe the odd behavior of the spin-orbit terms which
are generated as derivative of the central terms. The central terms
have quite a few oscillations and this generates big variations in
the spin-orbit term. We have seen in the previous section that the
results are not very sensitive to the details of the spin-orbit
term, however the shape of the potentials in the lower panel of
fig. 5 is quite odd.

To get rid of this problem we have substituted the spin-orbit term
given in eq. (2.4) as derivative of the central term, with a
Woods-Saxon term of the form (2.8). The three additional parameters
of the spin-orbit term are also entering in the fitting procedure.
The results of these fits are presented in tab. 2 and 3 where we have
called these results as SMHWS (Shell Model Hermite Woods-Saxon).

The central potentials of the Hermite model and of the Hermite
plus Woods-Saxon are compared in fig. 6. One can see that the
SMHWS potentials are smoother than the Hermite potentials, but in
general the shapes are quite similar.

The SMHWS potentials have reasonable behavior from the physics
point of view. We should remark that they are always negative and
they are smoothly going to zero for large values of the radius. The
only exception is the potential of $^{40}Ca$ which shows a strange
pocket around $9$ fm. All our attempts to get rid of this pocket
have produced fits of much worse quality.

The two mean field models produce fits of the same quality, as
well as it is possible to see from the results shown in tab. 2 and
3 and from fig. 4 where the charge distributions obtained with the
SMHWS approach are presented by the dashed lines.

The result obtained so far show that the mean field approach
produces fits of the same quality of those obtained using the
model independent techniques. In fig. 6 we present the densities
obtained by our fits in logarithmic scale. The densities produced by
the model independent techniques (bot with FB or Hermite expansion
basis) show clear anomalies in the tail, while the mean field
densities have the proper exponential decay compatible with the
quantum mechanics requirements.

\vskip 1 cm
{\bf 3. Conclusions.}
\vskip 0.5 cm

In this paper we have presented a method, alternative to the model
independent techniques, to extract the charge density distributions
from elastic electron scattering data.

The method is based on the mean field model of the nucleus. The
mean field potential is expanded on a orthonormal basis of Hermite
functions, and the expansion coefficients are modified in order to
reproduce the cross section.

Clearly this method is restricting the set of functions describing
the charge distribution, while model independent techniques are more
general. On the other hand, this restriction is stabilizing the
solution of the problem. We have shown in sect.2 and in fig. 3
that increasing the
number of coefficients creates oscillation in the mean field
potential, but these oscillations are not producing instabilities in
the charge distributions.

The loss in generality is slightly penalizing the quality of the
of the fit. In tab. 3 we observe that the $\chi^2$ values produced
by a model independent fit with a basis of Hermite functions are a
slightly better than the values obtained with our method.

The better performances of the model independent techniques have the
serious drawback that the charge distributions produced have an
unphysical behavior at high values of the nuclear radius.
On the contrary, the tails of the distributions
obtained with our method have the proper exponential behavior,
since they are produced by the solution of the Schr\"odinger
equation.
We have performed fits with model independent techniques forcing
the charge distributions to have an exponential tail after a large
value of the radius. The quality of these fits obtained by
imposing this physical restriction is analogous to
that obtained with our mean field approach.

Finally we want to remark that the mean fields we are generating have
only a pragmatical origin and aim. Any attempt to connect them to
microscopical theories of the nucleus is meaningless.

\vfill\eject

{\bf References}
\vskip 1cm

\item{[1]} K.Chadan and P.C.Sabatier, Inverse Problems in Quantum
Scattering Theory, \par
 (Springer, Berlin, 1977) p.423. \par
\vskip .5cm

\item{[2]} J.L.Friar and J.W.Negele, Nucl. Phys. A212 (1973)
93.\par
\vskip .5cm

\item{[3]} B.Dreher, J.Friedrich, K.Merle, H.Rothhaas and
G.L\"urs,  Nucl. Phys. A235 (1974) 219.\par
\vskip .5cm

\item{[4]} R.Anni, G.Co' and P.Pellegrino, preprint, Dip. di
Fisica, Univ. di Lecce (1994), to be published on Nucl. Phys.
\vskip .5cm

\item{[5]} L.R.Elton and A.Swift, Nucl. Phys. A94 (1967) 52.
\vskip .5cm

\item{[6]} D.R.Yennie, D.G.Ravenhall and R.N.Wilson, Phys. Rev.
95 (1954) 500.\par
\vskip .5cm

\item{[7]} A.Messiah, Mecanique Quantique, vol. 1
 (Dunod, Paris, 1962) p.418.\par
\vskip .5cm

\item{[8]} J.E.Amaro, G.Co' and A.M.Lallena, Ann. Phys. (NY)
221 (1993) 306. \par
\vskip .5cm

\item{[9]} G.G.Simon, Ch.Schmitt, F.Borkowski and V.H.Walther,
Nucl. Phys. A333 (1980) 381. \par
\vskip .5cm

\item {[10]} J.Heisenberg, R.Hofstadter, J.S.McCarthy, I.Sick,
B.C.Clark, R.Herman and \par
D.G.Ravenhall, Phys. Rev. Lett. 23 (1969) 152.\par
B.Frois, J.Bellicard, J.M.Cavedon, M.Huet, P.Leconte, A.Nakada,
P.X.Ho and I.Sick,\par
Phys. Rev. Lett. 38 (1977) 1259.\par
\vskip .5cm

\item {[11]} J.M.Cavedon, Th\`ese de doctorat d'Etat, Paris
1980, Unpublished.\par
\vskip .5cm

\item{[12]} A.Bohr and B.Mottelson, Nuclear Structure, vol. 1,
 (Benjamin, New York, 1969) p.218.\par
\vskip .5cm

\item{[13]} I.Sick and J.S.McCarthy, Nucl. Phys. A150 (1970)
631.\par
\vskip .5 cm

\item{[14]}
L.S.Cardman, J.W.Lightbody, S.Penner, S.P.Fivonzinsky
and X.K.Maruyama,\par
Phys. Lett. 91B (1980) 203.\par
W.Reuter, G.Fricke, K.Merle and H.Miska, Phys. Rev. C26 (1982)
806.\par
\vskip .5cm

 \item {[15]} B.B.P.Sinha, G.A.Peterson, R.R.Whitney, I.Sick and
J.S.McCarthy,\par
Phys. Rev. C7 (1973) 1930.\par
I.Sick, J.Bellicard, J.M.Cavedon, B.Frois, M.Huet, P.Leconte,
P.X. Ho and \par
S.Platchkov,
 Phys. Lett. 88B (1979) 245.\par
\vskip .5cm

\item {[16]} C.M.Lederer and V.S.Shirley, Table of Isotopes,
7th edition (Wiley, New York,  1978).\par
G.A.Rinker and J.Speth, Nucl. Phys. A306 (1978) 360. \par
G.Co', A.M.Lallena and T.W.Donnelly, Nucl. Phys. A469. (1987)
684.\par

\vfill\eject

\def\tablerule{\noalign{\hrule}}
\def\qq{\;\;}
\midinsert{
$$\vcenter {\offinterlineskip \hrule
\def\spal1{height5pt
&\omit&&\omit&&\omit&&\omit&&\omit&\cr}
\halign { & \vrule# &
          $\qq\hfil#\hfil\qq$ & \vrule# &
          $\qq\hfil#\hfil\qq$ & \vrule# &
          $\qq\hfil#\hfil\qq$ & \vrule# &
          $\qq\hfil#\hfil\qq$ & \vrule# &
          $\qq\hfil#\hfil\qq$ & \vrule# &
          $\qq\hfil#\hfil\qq$ & \vrule# &
          $\qq\hfil#\hfil\qq$ & \vrule# &
          $\qq\hfil#\hfil\qq$ & \vrule# \cr
\spal1
& \omit && V_o \,\,\, (MeV) && \lambda  && R \,\,\, (fm) &&
a \,\,\, (fm) & \cr \spal1
\tablerule \spal1
& \omit && \omit && \omit && \omit &\cr \spal1
& ^{208}Pb  && \ 60.0 && \ 0.3 && \ 7.5 &&  \ 0.6  & \cr\spal1
& \omit && \omit && \omit && \omit &\cr \spal1
}\hrule}$$
\vskip7pt \noindent
\baselineskip=.15in
{\hsize 9cm{\bf Table~1.}\ \
Parameters of the Woods-Saxon potential (eq. 2.8) producing the
density used to generate the pseudodata.
}}
\endinsert

\vfill\eject

\def\tablerule{\noalign{\hrule}}
\def\qq{\;\;}
\midinsert{
$$\vcenter {\offinterlineskip \hrule
\def\spal1{height5pt &\omit&& \omit &&\omit&&\omit&&\omit&\cr}
\halign { & \vrule# &
          $\qq\hfil#\hfil\qq$ & \vrule# &
          $\qq\hfil#\hfil\qq$ & \vrule# &
          $\qq\hfil#\hfil\qq$ & \vrule# &
          $\qq\hfil#\hfil\qq$ & \vrule# & $\qq\hfil#\hfil\qq$ &
          \vrule# & $\qq\hfil#\hfil\qq$ &\vrule# & $\qq\hfil#\hfil\qq$ &
          \vrule# \cr
\spal1
& Nucleus && \omit && SMH && SMHWS && exp. &\cr \spal1
\tablerule \spal1
& \omit && \omit && \omit && \omit && \omit &\cr \spal1
& ^{12}C && \ 1p3/2  && \ -15.53 && \ -16.00 &&  \ -15.96 &\cr \spal1
& \omit && \omit && \omit && \omit && \omit &\cr \spal1
& \omit && \ 1p1/2  && \omit && \ -3.5 && \ -1.94 &\cr \spal1
& \omit && \omit && \omit && \omit && \omit &\cr \spal1
\tablerule \spal1
& \omit && \omit && \omit && \omit &\cr \spal1
& ^{16}O && \ 1p3/2 && \ -24.17 && \ -18.45 && \ -18.44 &\cr \spal1
& \omit && \omit && \omit && \omit && \omit  &\cr \spal1
& \omit && \ 1p1/2  && \ -16.7 && \ -11.97 && \ -12.11 &\cr \spal1
& \omit && \omit && \omit && \omit && \omit \cr \spal1
\tablerule \spal1
& \omit && \omit && \omit && \omit && \omit &\cr \spal1
& ^{40}Ca && \ 1d5/2  && \ -12.02 && \ -15.75 && \ - 14.73 &\cr \spal1
& \omit && \omit && \omit  && \omit && \omit &\cr \spal1
& \omit && \ 2s1/2 && \ -8.67 && \ -10.71 && \ -10.33   &\cr \spal1
& \omit && \omit && \omit  && \omit && \omit &\cr \spal1
& \omit && \ 1d3/2 && \ -9.07 && \ -8.64 && \ -8.30   &\cr \spal1
& \omit && \omit && \omit && \omit && \omit \cr \spal1
\tablerule \spal1
& \omit && \omit && \omit && \omit && \omit &\cr \spal1
& ^{208}Pb && \ 1g7/2  && \ -11.88 && \ -12.78 && \ -11.51 &\cr \spal1
& \omit && \omit && \omit && \omit && \omit &\cr \spal1
& \omit && \ 2d5/2  && \ -9.96 && \ -10.40 && \ -9.71 &\cr \spal1
& \omit && \omit && \omit && \omit && \omit &\cr \spal1
& \omit && \ 1h11/2  && \ -9.10 && \ -9.22 && \ -9.37 &\cr \spal1
& \omit && \omit && \omit && \omit && \omit &\cr \spal1
& \omit && \ 2d3/2  && \ -8.36 && \ -8.88 && \ -8.39 &\cr \spal1
& \omit && \omit && \omit && \omit && \omit &\cr \spal1
& \omit && \ 3s1/2  && \ -7.65 && \ -8.12 && \ -8.03 &\cr \spal1
& \omit && \omit && \omit && \omit && \omit &\cr \spal1
& \omit && \ 1h9/2  && \ -4.21 && \ -5.21 && \ -3.80 &\cr \spal1
& \omit && \omit && \omit && \omit && \omit &\cr \spal1
& \omit && \ 2f7/2  && \ -3.37 && \ -3.81 && \ -2.90 &\cr \spal1
& \omit && \omit && \omit && \omit && \omit &\cr \spal1
& \omit && \ 1i13/2  && \ -2.45 && \ -2.50 && \ -2.19 &\cr \spal1
& \omit && \omit && \omit && \omit &\cr \spal1
}
\hrule}$$
\vskip7pt \noindent
\baselineskip=.15in
{\hsize 9cm{\bf Table~2.}\ \
Comparison with the experimental single particle energies, in MeV.
The experimental values have been taken from ref. [16].
The SMH values have been obtained with the Hermite expansion of the potential
presented in sect. 2. The SMHWS have been obtained expanding the central part
of
the potentiel on the Hermite function basis, and using a Woods-Saxon shaped
spin-orbit term. Note that in the SMH calculation the $1p1/2$ level of $^{12}C$
was unbound. }} \endinsert

\vfill\eject

\def\tablerule{\noalign{\hrule}}
\def\qq{\;\;}
\midinsert{
$$\vcenter {\offinterlineskip \hrule
\def\spal1{height5pt &\omit&&\omit&&\omit&&\omit&&\omit&\cr}
\halign { & \vrule# &
          $\qq\hfil#\hfil\qq$ & \vrule# &
          $\qq\hfil#\hfil\qq$ & \vrule# &
          $\qq\hfil#\hfil\qq$ & \vrule# &
          $\qq\hfil#\hfil\qq$ & \vrule# & $\qq\hfil#\hfil\qq$ &
          \vrule# & $\qq\hfil#\hfil\qq$ &\vrule# & $\qq\hfil#\hfil\qq$ &
          \vrule# \cr
\spal1
& Nucleus && \ SMH && \ SMHWS && \ FB && \ Hermite &\cr \spal1
\tablerule \spal1
& \omit && \omit && \omit && \omit && \omit &\cr \spal1
& ^{12}C && \ 1.42  && \ 1.25 && \ 1.73 && \ 1.11 &\cr \spal1
& \omit && \omit && \omit && \omit && \omit &\cr \spal1
\tablerule \spal1
& \omit && \omit && \omit && \omit && \omit &\cr \spal1
& ^{16}O && \ 4.44 && \ 4.01 && \ 3.55 && \ 3.38 &\cr \spal1
& \omit && \omit && \omit && \omit && \omit  &\cr \spal1
\tablerule \spal1
& \omit && \omit && \omit && \omit && \omit &\cr \spal1
& ^{40}Ca && \ 1.35  && \ 1.31 && \ 1.38 && \ 1.00 &\cr \spal1
& \omit && \omit && \omit  && \omit && \omit &\cr \spal1
\tablerule \spal1
& \omit && \omit && \omit && \omit && \omit &\cr \spal1
& ^{208}Pb && \ 1.59  && \ 1.52 && \ 1.44 && \ 1.39 &\cr \spal1
& \omit && \omit && \omit && \omit && \omit &\cr \spal1
}
\hrule}$$
\vskip7pt \noindent
\baselineskip=.15in
{\hsize 9cm{\bf Table~3.}\ \
Reduced $\chi^2$ for the fit to the experimental elastic cross sections
obtained
within our mean field model (SMH and SMHWS) and with the model independent
expansion techinques using a Fourier-Bessel (FB) and a Hermite basis.
 }} \endinsert

\vfill\eject

{\bf Figure captions}

\vskip .5cm

\item {Fig.1}
Results of the fit to the cross section pseudodata.
The upper panel shows the charge distributions, the middle panel the
central potentials and the lower panel the spin-orbit potentials. The
full lines show the quantities used to generate the pseudodata from
the Woods-Saxon potential. The dashed lines are generated fitting
the cross section with the potential constructed with eqs. (2.4) and
(2.5), and the dashed dotted lines by the model where the spin-orbit
potential is produced by the symmetric function around the first
maximum of the derivative of $V(r)$ starting from the external part
of the nucleus (see text).

\vskip .2cm

\item {Fig.2}
Charge distributions and central potential produced fitting the
pseudodata with various expansion terms of the potential, eq. (2.5).
\vskip .2cm

\item{Fig.3}
Charge distribution obtained fitting the $^{40}Ca$
data of ref. [11]. The shaded areas contains all the
charge distributions and potentials
obtained by a set
of 100 fits of the data normally sampled within the error band as
discussed in ref. [4]. The panels A and B shows the distribution
bands obtained with a Fourier Bessel expansion of the density. In A,
9 expansion coefficients have been used, and in B 13. In
the panel C and D the darker areas show the density bands obtained
with our method using 8 and 14 coefficients respectively. The
lighter areas, related to the right vertical scale,
show the dispersion bands of the potential. We add also the
information about the reduced averaged $\chi^2$ for each set of
fits.

\vskip .2cm

\item {Fig.4}
Results of the fit to the experimental data of
ref.[10,11,13,14,15]. The uncertainty bands are produced by the
Fourier-Bessel fit to the data, as discussed in ref. [4]. The full
lines, usually lying within the uncertainty band, are generated by
our mean field model. The dashed lines have been produced
substituting the model spin-orbit term with a Woods-Saxon one (see
text). \vskip .2cm

\item {Fig.5}
Central (upper panel) and spin-orbit (lower panel) potentials
produced by the fit to the experimental data. The full lines refer to
$^{12}$C, the dotted ones to $^{16}$O, the dashed ones to $^{40}$Ca
and the dashed dotted ones to $^{208}$Pb.
\vskip .2cm

\item {Fig.6}
Comparison between the central potentials of the previous
figure(full lines) and those generated fitting the cross section data
using a spin-orbit term of Woods-Saxon type (dashed lines).
\vskip .2cm

\item {Fig.7}
Charge distribution for the four nuclei considered presented in
logarithmic scale. The results of our model are shown by the full
and dotted lines. SMH indicates the model presented by the eqs.
(2.4) and (2.5) where the potential has been expanded on a Hermite
functions basis. SMHWS indicates the same model where the spin-orbit
term has been substituted with a Woods-Saxon term. The dashed and
dashed dotted lines have been obtained by a fit using the model
independent procedure where the density distributions have been
expanded on a Fourier-Bessel (FB) and Hermite basis respectively.
\vskip .2cm

\end